# Weedy Adaptation in *Setaria* spp.: VIII.  Structure of *Setaria faberi* Seed, Caryopsis and Embryo Germination

Milt Haar[1], Adriaan van Aelst[2] and Jack Dekker[1]

[1] Weed Biology Laboratory, Department of Agronomy, Iowa State University
Ames, Iowa 50011, USA
[2] Department of Biomolecular Sciences, Wagingen Agricultural University, Wageningen, The Netherlands

**Abstract**.  Giant foxtail *(Setaria faberi)* seeds differ in requirements for germination.  Variable germinability arises during seed development under the influence of genotype, environment and parent plant. Giant foxtail seed germination has been shown to be regulated by independent asynchronous or dependent synchronous action of seed structures. To gain better insight into the process, germination was divided into axis specific embryo growth categories or states. Three states were defined for each embryonic axis. The degree of embryo growth (germination state) after eight to twelve days under germination conditions is believed to reveal the germinability state (potential for germination) possessed by the seed before germination. The embryo axes behave independently, which allows any combination of germination states to occur. In general, the greater the difference in germination between the axes, the less likely the combination of states will occur.  Photographic evidence of each germination state is shown for caryopses and seed.  Seed with a variety of germinability states is a strategy for surviving variable environments.

## INTRODUCTION

Giant foxtail is a major weed of the U. S. cornbelt (Holm et. al., 1977; Knake, 1977). Seed germinability (dormancy) is one of the primary factors responsible for the success of foxtails and other grasses as weeds (Simpson, 1990). The addition of dormant seed to the seed bank serves to disperse foxtail over time and enables the species to survive unfavorable environmental conditions or weed control efforts (Cavers, 1995; Grime, 1981).  The annual seed rain of a foxtail plant consists of individuals with different germination requirements, each seed shed from the parent plant having potentially a different germinability state (Dekker, et. al., 1996; Trevewas, 1987).  Shedding heterogeneous seed with different germinability states is a well-adapted strategy for survival in highly variable environments (Silvertown, 1984). Germinability states are believed to arise during seed development under the influence of the parent plant and the environment (Bewley and Black, 1994; Simpson, 1990). The degree of germinability is not fixed after embryogenesis and may change in response to environmental conditions at any time during the seed phase of the plant life cycle. (Trevewas, 1987).

There is evidence that embryo germination in foxtail spp. is influenced by other seed structures.  The dispersal unit for giant foxtail is referred to as a "seed", although the term, spikelet, is more accurate. Each seed consists of a single floret subtended by a sterile lemma and two glumes (Narayanaswami, 1956; Rost, 1975). A lemma and palea surround the caryopsis and together form a hard protective covering commonly referred to as the "hull" (Gould, 1968; Rost, 1975) (Figure 1; see also figure 10, lower). Physical removal of the hull has been observed to stimulate germination of foxtail seeds (Biswas et. al., 1970; Martin, 1943; Haar and Dekker, 1998; Kollman, 1970; Nieto-Hatem, 1963; Povilaitis, 1956; Rost, 1975). Hull damaging treatments such as scarification or piercing increased germination (King, 1952; Kollman and Staniforth, 1972; Peters et. al., 1963; Stanway, 1971).

**Figure 1**. The seed hull of *Setaria faberi*; top: interior cross-section, surface; bottom: surface topography. (Photo: Haar-vanAelst-Dekker)

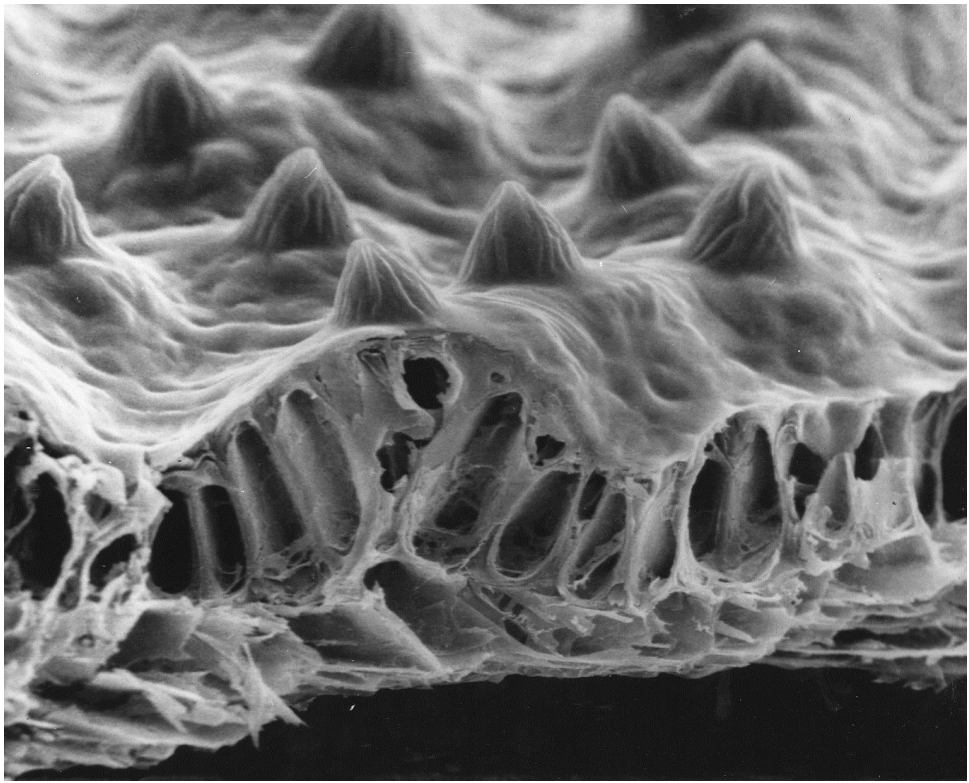

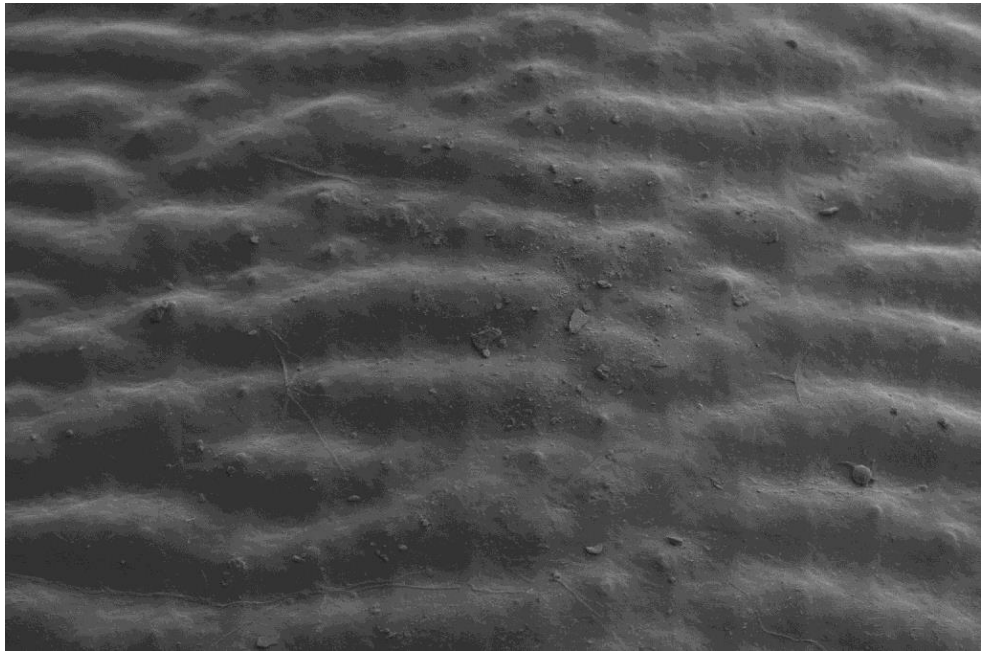

Attempts to explain the mechanism by which foxtail spp. hulls affect germinability have focused on three theories: 1) physical restraint of the hull on the embryo 2) presence of an inhibitory substance and 3) impermeability to water and gases. Hulls do not appear to significantly impede water absorption in the foxtail spp. Germinable and dormant foxtail seeds imbibed similar amounts of water (Kollman, 1970; Neito-Hatem, 1963). Evidence for a chemical inhibitor is inconclusive; leaching increased percent germination of intact and damaged seeds of yellow foxtail; hull leachate, however, failed to inhibit caryopses germination (Kollman, 1970; Nieto-Hatem, 1963). Nieto-Hatem (1963) described two levels of yellow foxtail seed germinability: hull and caryopsis. Removal of the hull allowed some caryopses to germinate, these seeds were defined as having hull germinability. Other caryopses did not germinate when hulls were removed and were regarded as caryopsis germinable. Interior to the hull is the caryopsis, composed of the embryo, endosperm and caryopsis coat. The surface of the caryopsis is a layer 3-10 um thick known as the caryopsis coat. Several compressed tissue layers including the integuments, nucellus and pericarp make up the caryopsis coat (Rost, 1973). The caryopsis coat has been implicated in affecting seed germinability; usually as an impediment to gas or water movement. It has been observed that disruption of the caryopsis coat stimulates yellow foxtail embryos to germinate (Rost, 1973). Excision of embryos from dormant seeds also stimulates their germination (Rost, 1971 and 1975; Dekker et. al., 1996) (Figure 2).

**Figure 2**. The embryo and scuttelum (sct) of *Setaria faberi*; embryo axes: coleoptile (ct) and coleorhiza (cr). (Photo: Haar-vanAelst-Dekker)

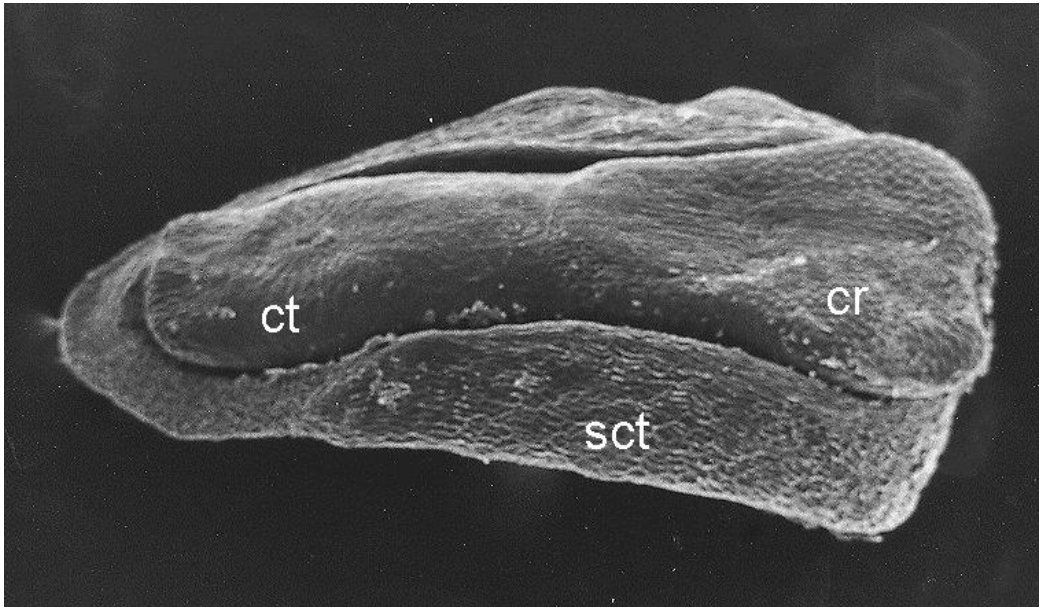

There is no single, qualitative state of germinability into which all foxtail seeds must enter, remain, and exit (Simpson, 1990). Instead there are many germinability states among different seed shed by a parent plant arising from the wide array of environmental conditions that parent plant experiences. These heterogeneous conditions induce developmental arrest (dormancy) to different degrees in different seeds as they develop as a function of changing sensitivities of seed organs from anthesis through after-ripening. Foxtail seed rain consists of these diverse germinability phenotypes as a consequence of influences exerted at the level of plant tiller, inflorescence, fascicle, spikelet, and seed compartments (Dekker et al., 1996, Haar and Dekker, 1998).

     A previous paper in this series or articles on foxtail spp. weedy adaptation provided evidence for a complex model of germinability regulation based on the independent, asynchronous actions of the embryo, caryopsis and hull compartments, as well as on their dependent, synchronous action (Dekker et. al., 1996). Embryo germination can take place in either or both axes (coleoptile, shoot; coleorhiza, root), axis-specific germination. The germinability of isolated embryos is not necessarily an indication of their germinability when enclosed in the caryopsis or hull, often surrounding envelopes inhibit germination ((Dekker et. al., 1996); Haar and Dekker, 1998).

     The goal of this study is to provide photographic evidence of the several morphological germination states of giant foxtail quantitatively described previously (Dekker et al., 1996). We provide herein micrographic presentations of the continuous nature of giant foxtail embryo germination process. The germination process for this study was broadly defined and divided into categories or states based on this continuous embryo growth. It is believed that more insight into the behavior of seed germination can be gained by using several quantitative states of embryo growth. It may also provide additional insights into the nature of foxtail spp. seed dormancy and seed bank dynamics.

## Materials and Methods

**Plant Material**. A single seed was used to propagate the seed used in these studies (single seed descent propagation scheme; our seed lot number 1816; Dekker et al., 1996) in order to reduce variation among observations due to differences in genotype. Seed was harvested in 1991 and stored dry at 4 to 6°C in darkness until used in this study. Only dark brown, fully mature seeds were in used. Germination was greater than 90 percent at the time this study took place. The hull was removed with a scalpel and forceps under a dissecting microscope for observations of caryopses and embryos (Dekker et al., 1996).

**Germination**. Prior to germination seed was surface sterilized by immersion in a ten percent commercial bleach solution for 15 minutes then rinsed with distilled water. Seeds or caryopses were placed in 5.5 cm diameter glass petri dishes on two five cm filter paper (No. 595; Schleicher and Schuell, Dassel, Germany) discs moistened with 1.5 ml distilled water. Petri dishes were sealed with Parafilm"M" (American National Can, Neenah, WI 54956) then placed in a controlled environment chamber at 25°C and constant light.

**Electron microscopy**. Seeds or caryopses were observed at various stages of germination under one of two electron microscopes at the Wageningen Agricultural University, Wageningen The Netherlands. For observations under field scanning electron microscope (JSM 6300 F, Jeol). Specimens were mounted using tissue tech (source), frozen in supercooled liquid Nitrogen, then warmed to -85°C at 7 torr. Samples were then sputter coated with gold palladium.

Other specimens were fixed in four percent formalin in a phosphate buffer solution (pH 7.4) followed by dehydration in an ethanol series. Specimens were then critical point dried in carbon dioxide, mounted on Aluminum studs with double sided tape and sputter coated with gold palladium for two minutes. Observations were made using a scanning electron microscope.

## Results and Discussion

The process of germination for seeds and caryopses consists of a continuous chain of events which we have for descriptive and comparative purposes divided into several discrete stages (Figure 3). Three germination states were defined for each embryonic axis. An example of each state is shown.

**Caryopsis and Embryo Germination**

Isolated caryopses permit observation of the early embryo germination events that in seed are covered by the hull. The embryo is approximately one half the caryopsis length and is found near the surface of the caryopsis covered only by the caryopsis coat. The embryo scuttelum surrounds the embryonic axis below and along the margins, forming a cuplike structure in which the axis lies (Rost, 1973). Before imbibition the caryopsis is dry, the caryopsis coat wrinkled and the embryo sunken within the endosperm. This germination state is referred to as N (no germination; Figures 3a and 4). As the caryopsis imbibes water the embryo swells and longitudinal cracks begin to appear in the caryopses coat over the middle of the embryo axis (Figure 5). The cracking of the caryopsis coat due to growth defines the S1 state in caryopses (S for shoot apex; Figure 3b). Interpretation of the S1 state in isolated embryos was categorized by extension of the coleoptile beyond the scuttelum. It is possible, if S1 was the final germination state the individual seed reached, that the state could be due to water imbibition alone. Only cellular or molecular analyses may reveal whether S1 caryopsis coat cracking was due to germination growth or imbibition.

**Figure 3**. Giant foxtail germination states. Germination states are embryo axis specific and defined by growth: a) N, no germination; b) SlR0, first shoot (S1) growth state, embryo swells and cracks appear in caryopsis coat; c) S2R0, coleoptile extends and rises above caryopsis; d) S3R0, cotyledon emerges from coleoptile; e) first root axis growth state (Rl), col eorhiza expands; f) S0R2, trichomes form on coleorhiza; g) S0R3, radicle emerges from coleorhiza.

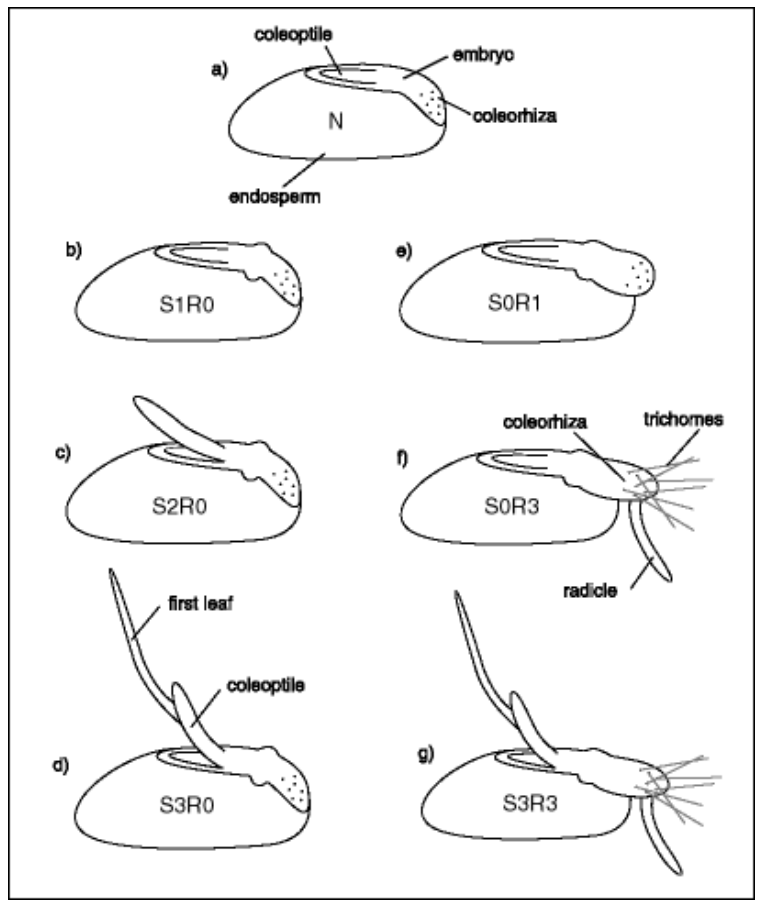

In most observations cracking and embryo swelling in shoot tissues of S1's were sufficient to eliminate imbibition as the sole explanation for the increase in embryo size and caryopsis coat cracking. It was also noted that occasionally upon dissection of intact ungerminated seed, arrested S1 caryopses were found within.

**Figure 4**. The dry shrunken embryo axis (ea) within the endosperm of a giant foxtail caryopsis. Magnification = X80. (Photo: Haar-vanAelst-Dekker)

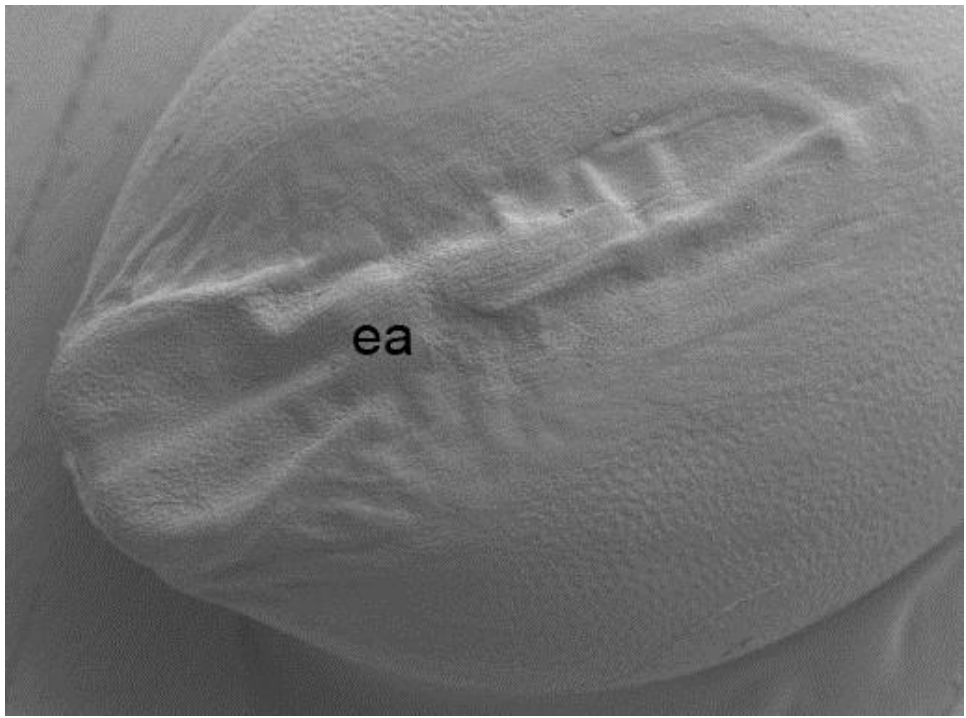

After imbibition, growth begins in germinable caryopses and the embryo increases in length. The second stage of germination for the shoot apex, S2, is determined by the upward bending of the coleoptile (Figures 3c and 6) followed by the emergence of the first true leaf, the cotyledon, from the coleoptile, which is the third stage, S3, in shoot germination (Figures 3d and 7).

**Figure 5**. S1R1 germination state. Giant foxtail caryopsis with swollen embryo. Longitudinal cracks (arrows; ↓) in the caryopsis coat appear above the embryo. Magnification = X50. (Photo: Haar-vanAelst-Dekker)

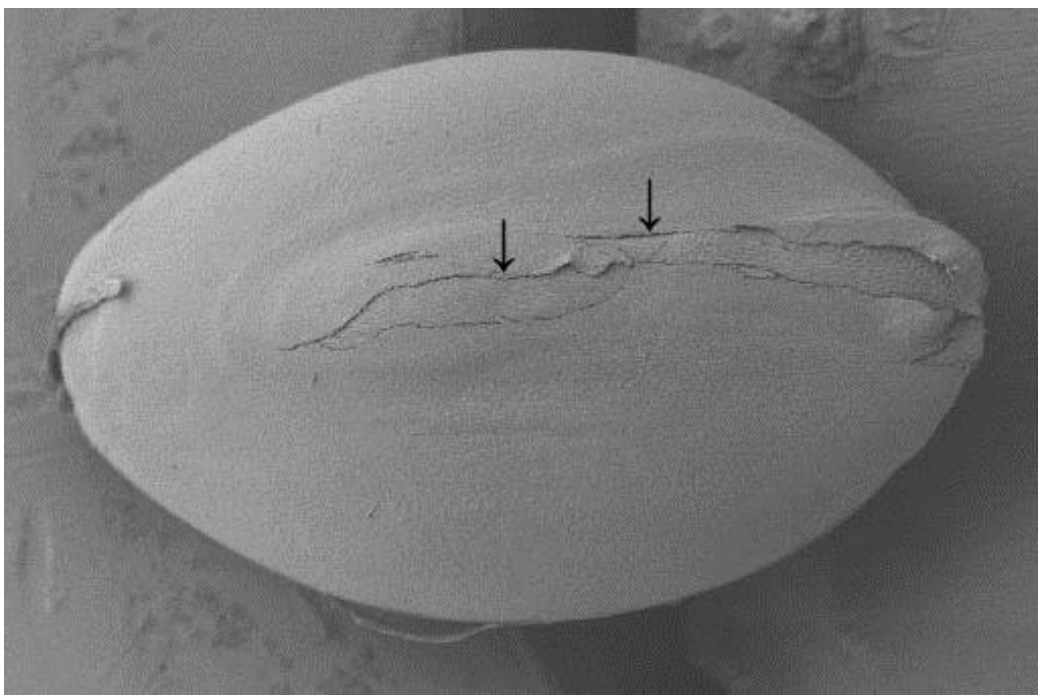

**Figure 6**. S2R3 germination state; alternate views. The coleoptile (cp) is curving upward while the coleorhiza is covered with trichomes (t) and the radicle (r) has emerged. Magnification = X35. (Photo: Haar-vanAelst-Dekker)

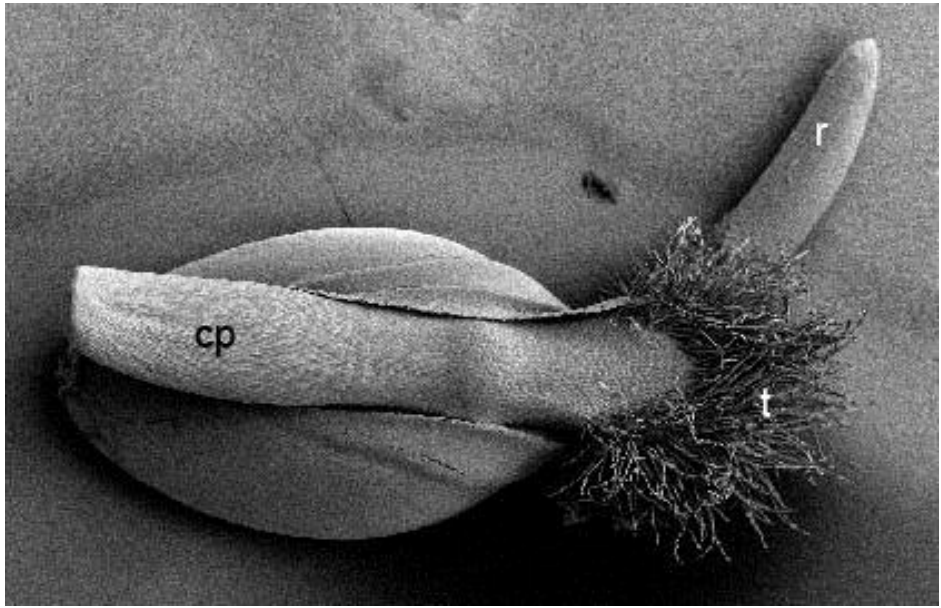

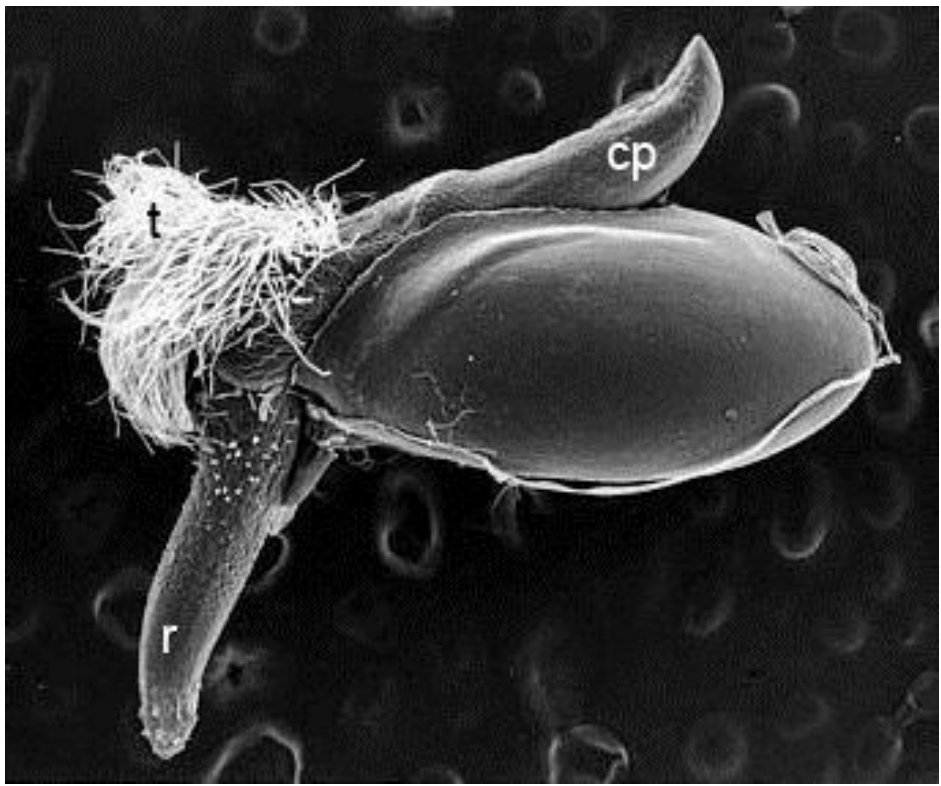

**Figure 7**. S3 germination state.  First true leaf, cotyledon (ct), emerges from coleoptile (cp). Magnification X50. (Photo: Haar-vanAelst-Dekker)

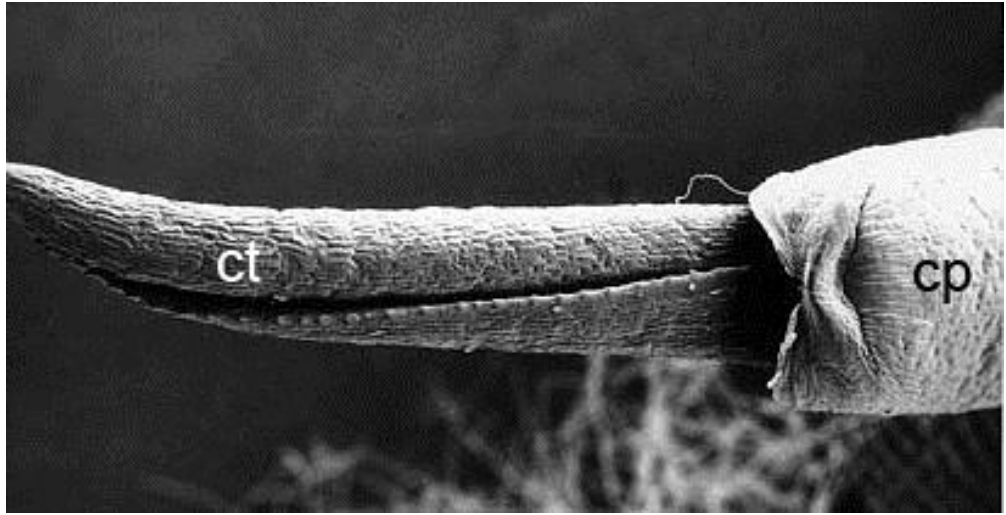

    The growth of the coleorhiza is the first germination event to occur for thefor the root axis. When the coleorhiza has extended beyond the scuttelar cup, the R1 (R for root axis) germination stage has been achieved (Figures 3e and 8). Initiation of trichomes on the coleorhiza advances germination to stage R2 (Figure 9). Trichomes begin to appear at the coleorhiza margins then expand to cover the coleorhiza (Figure 6). The observed degree of trichome growth was variable and in some cases absent. Water absorption and seedling anchoring are thought to be trichome functions (Northam et al., 1996). The third and final embryo germination state for the root axis, R3, is emergence of the radicle from the coleorhiza (Figures 3f; 6).

**Figure 8**.  SlRl germination state; alternate views.  Coleorhiza (cr) and coleoptile (cp) expanding.  Arrow (↓) points to crack in caryopsis coat. Magnification = X 65.  (Photo: Haar-vanAelst-Dekker)

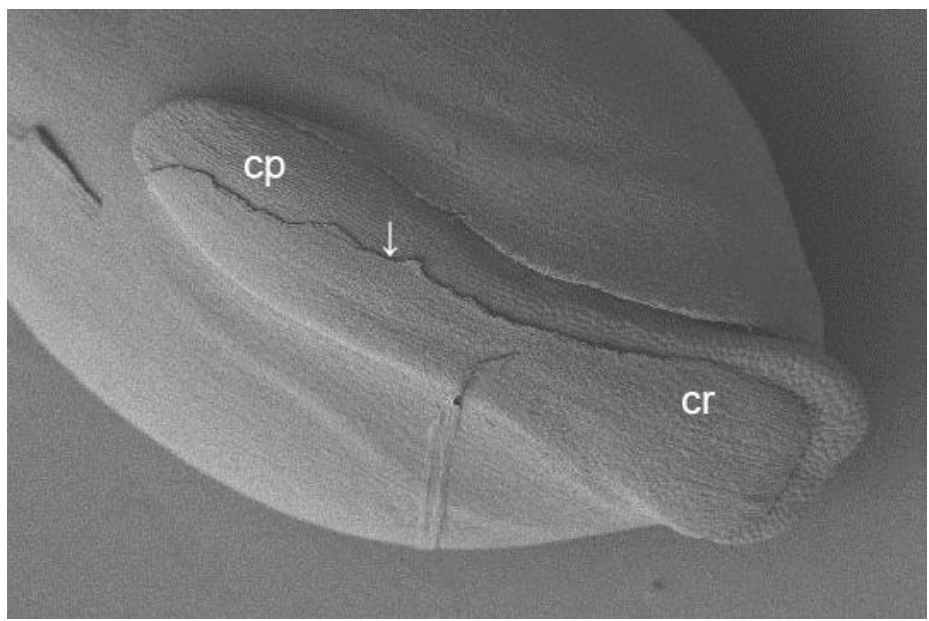

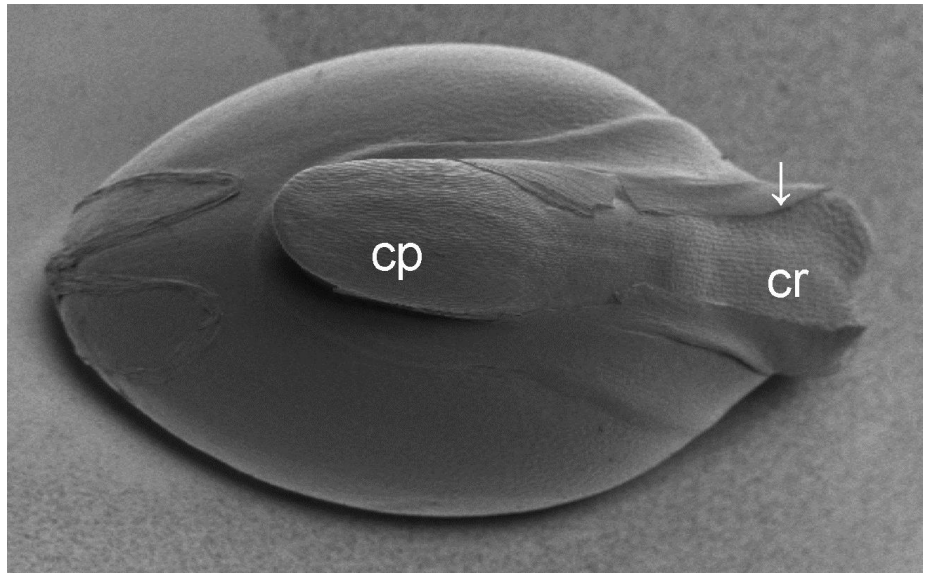

**Figure 9**. S1R2 caryopsis germination state; alternate views. Trichomes (t) are found at the margins of the coleorhiza. Magnification = X100. (Photo: Haar-vanAelst-Dekker)

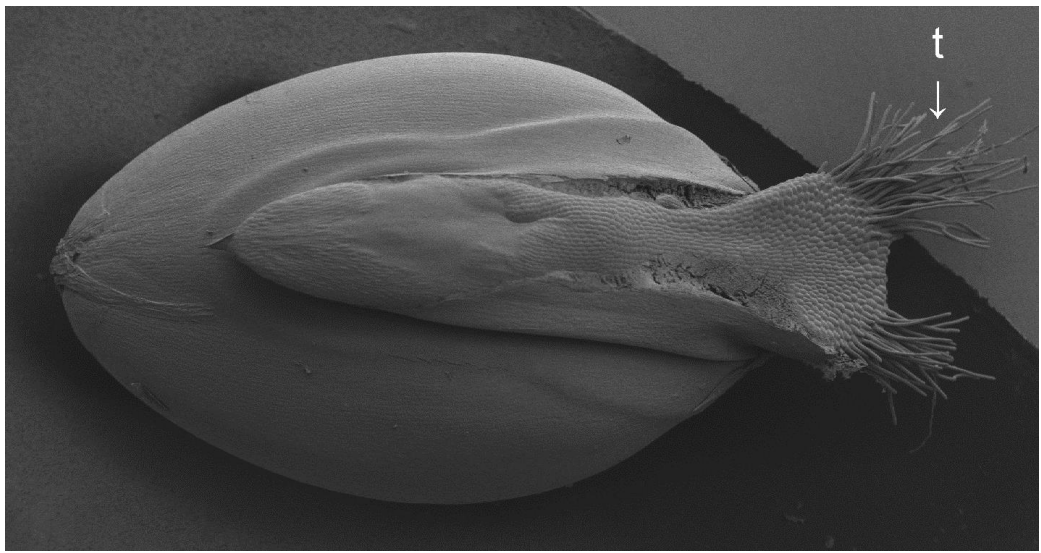

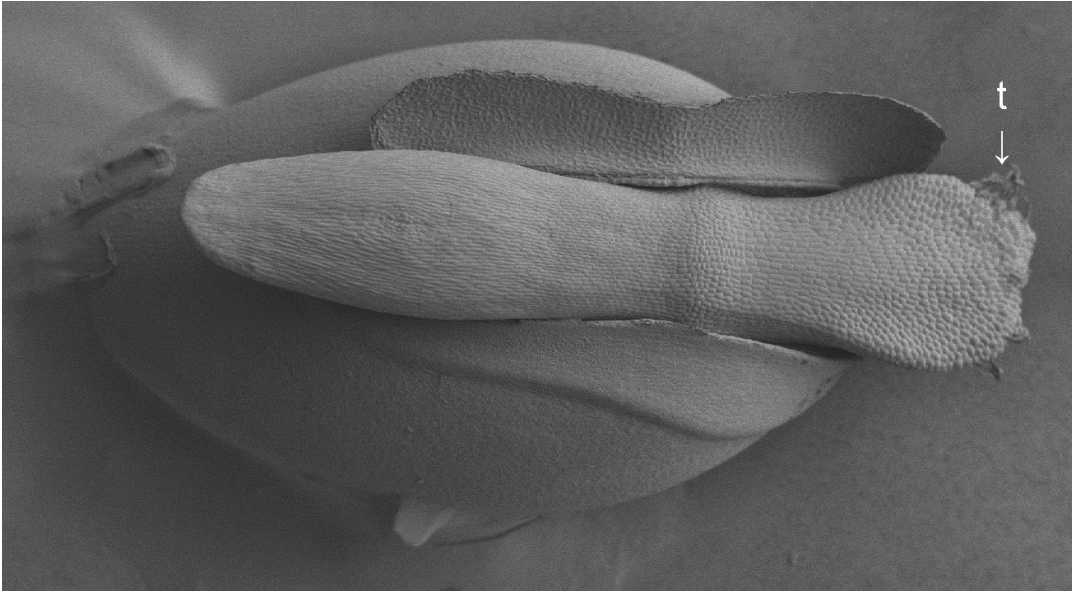

**Seed Germination**

A door or lid-like structure known as the germination lid is found at the proximal end of the lemma. It is through this structure that the coleorhiza exits the hull during germination (Rost, 1975). The hull exists in one of two states depending on whether the germination lid is open or closed (Dekker et al., 1995). The germination lid is attached hinged to the lemma on one side; the other sides abut against the lemma. We observed in these investigations that the three unhinged sides lack any physical connection to the adjacent lemma. The hinge provides the only resistance to the opening of the germination lid which implies a physical mechanism for hull dormancy. The first evidence of foxtail seed germination is usually protrusion of the coleorhiza through the germination lid (germination state R1 for seed, Figure 10). Other root axis states are the same as those described for caryopsis germination. Concurrent with coleorhiza and radicle growth is growth of the other embryo axis. Opening of the lemma and palea at the apical end of the seed is the first indication of shoot growth, followed by the emergence of the coleoptile (germination state S2 for seed; Figure 11). Initially the direction of coleoptile growth is controlled by the shape of the lemma, but soon orientation is upward. After a period of growth the first true leaf emerges from the coleoptile and the S3 state is achieved.

**Figure 10**. Germination lid (gl). Upper: side view, slightly open and the colerhiza (cr) emerging; magnification = X150. Lower: basal end of seed view. (Photo: Haar-vanAelst-Dekker)

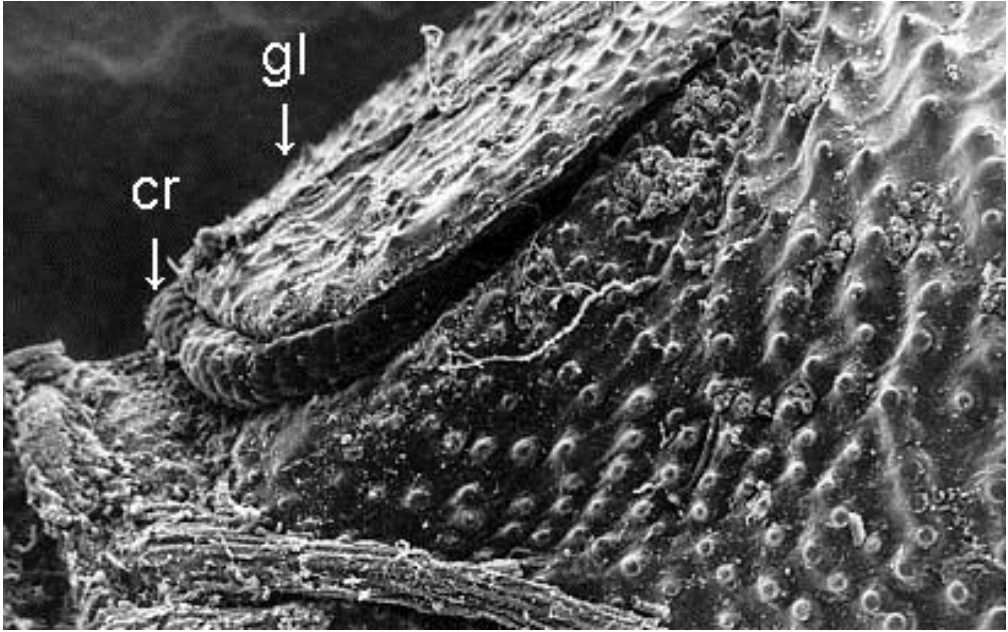
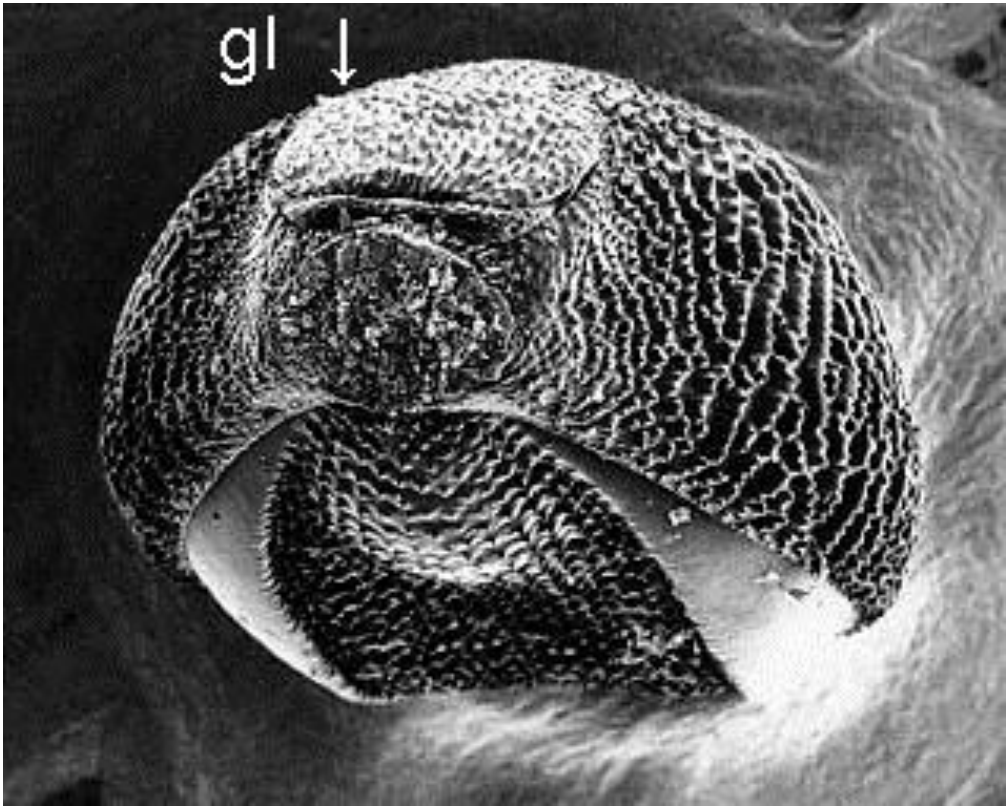

**Figure 11**. Seedling axis emergence. Top: coleoptile (cp) from the distal end of the seed; lemma (le), above; palea (pa), below; magnification = X100. Bottom: coleorhiza from the basal end germination lid (gl) (Photo: Haar-vanAelst-Dekker)

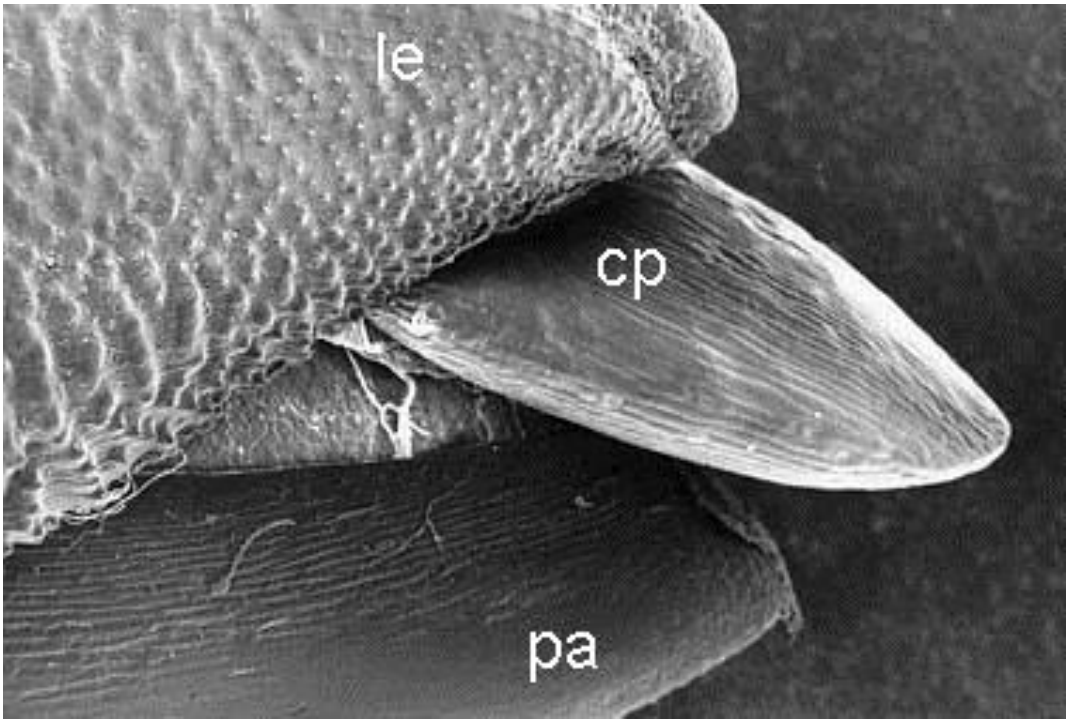

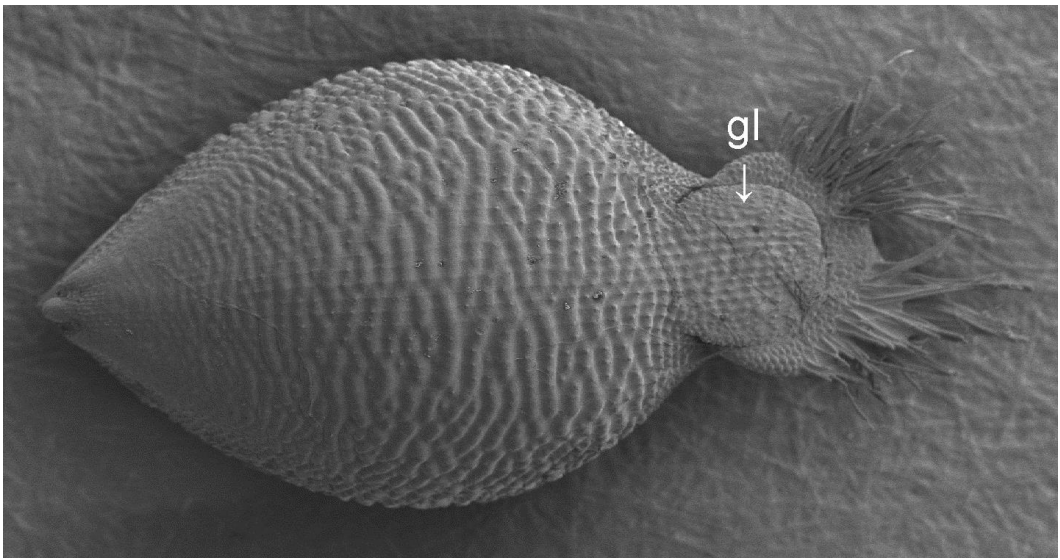

**Relative Sequence of Events**

Either apex must follow a sequential pattern of development (i.e. R3 cannot precede R1). The most typical germination state of giant foxtail seed is when germination takes place in both axes so that a seed, caryopsis, or embryo is described by the germination stage of both its root and shoot, e.g. S1R2. Typically in excised caryopses and embryos the shoot axis will germinate somewhat sooner than that in the root axis, although there is considerable variation in the relative timing of axis specific germination within one embryo. In seed, evidence of germination is first apparent by the emergence of the coleorhiza from the germination lid, and only rarely does the emergence of the coleoptile from the distal end occur first.

The two axes behave independently of one another and this independent axis growth makes possible several combinations of embryo germination: coleoptile only, coleorhiza only, both axes, and absence of axis germination (Dekker et. al., 1996). Some combinations occur more frequently than others. In general, the greater the relative difference in germination between the axes the less likely the combination will occur. It is thought that the variety of embryo axis combinations of growth patterns are the result of differences in the amount of developmental arrest induced in axis tissues during embryogenesis, and maintained in those tissues subsequent to abscission. This axis-specific germinability may be an important contribution to the production of heterogeneous seed and heterogeneous behavior of that seed in the soil seed bank (Forcella et al., 1992, 1997). Variable seed rain germinability allows a plant to disperse seed through time, and the behavioral sensitivity of individual seeds allows the population to respond to a heterogeneous environment (Skelly, 1996; Trevewas, 1987).